\documentclass[preprint,pra]{revtex4}
\usepackage{amssymb,amsmath,graphicx,amscd,xcolor}
\usepackage{float}
\begin{document}

\title{Vacuum Radiation Pressure Fluctuations on Atoms}

\author{L. H. Ford}
\email{ford@cosmos.phy.tufts.edu}
\affiliation{Institute of Cosmology, Department of Physics and Astronomy,
Tufts University, Medford, Massachusetts 02155, USA}

\begin{abstract}
Recent work has shown that the stress tensor components, such as energy density or pressure, of a quantum field can
be subject to large vacuum fluctuations. The energy density or pressure must be averaged in time before the fluctuations 
can be finite, and the probability of a large fluctuation depends upon the details of the averaging and can be much larger
than that predicted by a Gaussian distribution. This paper explores vacuum radiation pressure fluctuation on Rydberg
atoms and their possible observable effects. The excitation and de-excitation of a Rydberg atom provide an explicit
model for the time averaging of the radiation pressure, as the atomic polarizability becomes time dependent, first increasing
and then decreasing again by several orders of magnitude. This switched polarizability can induce large vacuum 
pressure fluctuations, which can in turn temporarily transfer linear momentum to the atom and cause a recoil which might
be observable. 
\end{abstract}

\maketitle
\baselineskip=14pt	

\section{Introduction}
\label{sec:intro}

Recent work has shown that components of the stress tensor for a quantized field can exhibit large vacuum 
 fluctuations~\cite{FewsterFordRoman2010,FFR2012,FF2015,SFF18,FF20,WSF21}. The relevant quantum operator must be averaged in time
 or in space and time, and the resulting probability distribution is sensitive to the details of this averaging, but typically
 decreases as an exponential of a small fractional power. Hence the probability of a large vacuum fluctuation can be
 much larger than would be predicted by the Gaussian distribution which governs random processes. This reflects the
 fact that vacuum fluctuations exhibit strong correlations. The averaging in time, or space and time, may be viewed as arising from
 the details of the measurement of the stress tensor. 
 
 Quantum stress tensor fluctuations can lead to passive fluctuations of the gravitational field, and are hence a type of
 quantum gravity effect which might play a role in the early universe~\cite{FMNWW10,WHFN11}. Analogous effects
 might be observable in condensed matter system, such a quantum density fluctuations in a fluid~\cite{WF20}.

 In this paper, a specific model for the measurement of vacuum radiation pressure on an atom will be discussed. It
 involves a Rydberg atom which is excited to a high energy level by a short laser pulse, and later de-excited by another pulse.
 During this time, the polarizability of the atom can increase by many orders of magnitude before decreasing again. This
 time dependent  polarizability  acts to average the radiation pressure operator in time~\cite{HF17}  resulting in potentially 
 observable velocity fluctuations of the atom.
 
 The outline of this paper is as follows: We first review previous results on the fluctuations of linear quantum field
 operators is Sec.~\ref{sec:linear}, and then discuss the significance of measurements in finite intervals in Sec.~\ref{sec:compact}.
 The resulting probability distributions for the stress tensor operators are reviewed in Sec.~\ref{sec:stress}. In Sec.~\ref{sec:2time}, we introduce an
 averaging function depending upon two time scales which will be used later in the paper. The dependence of the probability distribution
 upon the details of the averaging is discussed in Sec.~\ref{sec:eta-beta}, and in Sec.~\ref{sec:atoms}  we review the case of radiation pressure
 fluctuations on atoms. In Sec.~\ref{sec:v-flucts}, our model for switching the polarizability of an atom is presented, and estimates for both the expected
 atomic speed fluctuations and the probability of large fluctuations are presented. The results of the paper are summarized in Sec.~\ref{sec:final}.
 
Lorentz-Heaviside units in which $\hbar = c =1$ are used throughout the paper, except as otherwise noted.

\section{Vacuum Fluctuations of Quantum Field Operators}
\label{sec:t-ave}

\subsection{The Case of Linear Field Operators}
\label{sec:linear}

It is well known that the vacuum fluctuations of a linear field operator, such as the electric field, satisfy a Gaussian probabilitiy
distribution. However the operator must first be averaged in either time or space with a test or sampling function. The use of
test functions has long been used in rigorous approaches to quantum field theory as a formal tool to obtain well-defined 
operators~\cite{PCT}. However, we may also view these functions as describing the physical effects of a measuring apparatus.
Let $f(t)$ denote a temporal sampling function which vanishes as $t \rightarrow \pm \infty$ and is normalized so that
\begin{equation}
\int_{-\infty}^\infty f(t) \, dt = 1\,.
\label{eq:f-norm}
\end{equation}
Now let $E({\bf x},t)$ be one cartesian component of the free electric field operator, and define its time average by
\begin{equation}
{\bar E} = \int_{-\infty}^\infty f(t) \, E({\bf x},t)  \, dt \,.
\label{eq:E-ave}
\end{equation}
Now ${\bar E}$ will undergo Gaussian fluctuations around a mean value of $\langle 0| {\bar E}|0\rangle = 0$ with a finite variance of
\begin{equation}
\sigma =  \langle 0| {\bar E}^2 |0\rangle  \,.
\label{eq:sigma}
\end{equation}
If $\tau$ is the characteristic duration of the sampling, then $\sigma \propto \tau^{-4}$, with the constant of proportionality dependent upon
the detailed functional form of $f(t)$. As $\tau \rightarrow 0$, the variance grows, $\sigma \rightarrow \infty$. The mathematical reason for 
this is that smaller values of $\tau$ cause higher frequency modes to give the dominant contribution to ${\bar E}$. The physical reason
is that measurements on shorter time scales probe larger vacuum fluctuations.
Here we have concentrated on time averaging, but averaging in space alone or in both space and time produce similar results for linear field operators.

The vacuum fluctuations of the electric field can have observable effects. Welton~\cite{Welton} showed that the dominant contribution 
to the Lamb  shift may be estimated by a simple argument involving the effects of electric field fluctuations on energy levels of the hydrogen 
atom. It was argued in Ref.~\cite{HF15} that vacuum electric field fluctuations will slightly increase the probability  of barrier penetration
by a charged particle, and that a simple estimate using the time averaged electric field operator agrees with a one-loop perturbation
theory calculation~\cite{FZ99}.

\subsection{Measurements in a Finite Interval}
\label{sec:compact}

It is often convenient to select simple functions, such as  Lorentzians or Gaussians for the sampling functions. However, these have tails 
which strictly correspond to a measurement which began in the infinite past and extends to the infinite future. A better choice is a function
which is strictly zero outside of a finite interval, a function with compact support. Such a function is necessarily non-analytic, but can be
taken to have all of its derivative finite, a $C^\infty$ function.  Some examples of compactly supported functions are constructed and 
discussed in Refs.~\cite{FF2015,FF20}. Here we consider such a function of time, $f(t)$, with duration of the order of $\tau$, the sampling time.

A key feature of compactly supported functions is the rate at which their Fourier transforms decay. Let the Fourier transform of $f(t)$ be
\begin{equation}
\hat{f}(\omega) =\int_{-\infty}^\infty dt \, {\rm e}^{-i\omega t}\, f(t) \,.
\label{eq:Fourier}
\end{equation}
If $f(t)$ is a compactly supported $C^\infty$ function, then $\hat{f}(\omega)$ will fall more rapidly that any power of $\omega$, but more
slowly than an exponential function as $|\omega| \rightarrow \infty$. 
Here we take both $f(t)$ and $\hat{f}(\omega)$ to be real, even functions, and consider a class of compactly supported functions for
which
\begin{equation}
\hat{f}(\omega) \sim \gamma \,  {\rm e}^{- \beta |\omega \tau|^\eta}, \qquad |\omega|\to\infty
\label{eq:fasympt}
\end{equation}
for some constants $\gamma$, $\beta >0$, and $0< \eta <1$. (Note that the parameter $\eta$ is denoted by $\alpha$ in Refs.~\cite{FF2015,FF20}, but in this
paper we reserve $\alpha$  to denote atomic polarizability.) In all cases, $\hat{f}(\omega)$ decreases more slowly than exponentially for large
$ |\omega|$.  The value of $\eta = 1/2$ is of physical
interest. A simple electrical circuit is given in Ref.~\cite{FF2015} in which the current increases as a function of time with $\eta = 1/2$ after
a switch is closed. 

The value of $\eta$ is linked to the rate of switch-on and switch-off of  $f(t)$; the more sudden the switching, the smaller will be the value of $\eta$,
and the more important will be the contribution of high frequency modes. This contribution will play an important role in the probability of
large vacuum fluctuations for quadratic operators.

\subsection{Probability Distributions for Stress Tensor Operators}
\label{sec:stress}

The case of operators which are quadratic in the fields, such the energy density and other components of the stress tensor is more 
complicated, and has been treated in several recent papers~\cite{FewsterFordRoman2010,FFR2012,FF2015,SFF18,FF20}.
Here the operator must be averaged in time in order to have finite fluctuations. Additional averaging in space is optional and is
discussed in Ref.~\cite{FF20}. Let $\bar{T}$ be the time average of a normal ordered quadratic operator $T(t)$, defined as
in Eq.~\eqref{eq:E-ave}. This averaged quadratic operator may be expanded in normal modes as 
\begin{equation}
\bar{T} = \sum_{i\, j} (A_{i j}\, a^\dagger_i \,a_j + B_{i j}\, a_i \,a_j + 
B^*_{i j} \, a^\dagger_i \,a^\dagger_j ) \,.
\label{eq:T}
\end{equation}
Here $a^\dagger_i$ and  $a_i$ are the creation and annihilation operators for mode $i$ of a bosonic field. Quantization in a finite
volume is assumed, so the modes are discrete. The $A_{i j}$ and $B_{i j}$ are symmetric matrices which are proportional to
$\hat{f}$. Specifically, we expect
\begin{equation}
A_{i j} \propto (\omega_i\, \omega_j)^{p/2 -1} \; \hat{f}(\omega_i - \omega_j) \,, \qquad 
B_{i j} \propto (\omega_i\, \omega_j)^{p/2 -1} \; \hat{f}(\omega_i + \omega_j) \,,
\end{equation}
where $p$ is an odd integer which depends upon the dimensions of the operator $\bar{T}$. For example, for $T = \varphi^2$, the square
of a massless scalar field , $p=1$, For $T = \dot{\varphi}^2$, or a typical component of the stress tensor for a massless field,  $p=3$. 
In Sec.~\ref{sec:atoms}, we will consider operators with higher time derivatives, and hence larger values of $p$.

One approach to finding the probability distribution for the vacuum fluctuations of $\bar{T}$ is to study the behavior of the moments,
\begin{equation}
\mu_n = \langle \bar{T}^n \rangle \,,
\end{equation}
where the expectation value is in the vacuum state. Because $\bar{T}$ is normal ordered, the first moment vanishes, $\mu_1 = 0$. The
probability distribution has equal area on either side of  $\bar{T} = 0$, so a given measurement is equally likely to return a negative value
as a positive value, even for quantities such as energy density which are non-negative in classical physics. This does not mean that the
distribution is symmetric. For quantities such as the energy density, the odd moments are non-zero and the distribution is skewed.
The width of the distribution 
is described by the second moment, or variance,
\begin{equation}
\mu_2 = 2 \sum_{i\, j} B_{i j}\, B^*_{j i} \, ,
\label{eq:m2}
\end{equation}
The fact that $\omega_i + \omega_j \geq 0$, combined with the decrease of $\hat{f}$ for increasing values of its argument, guarantees that
the variance is finite so long as a time average over a nonzero interval has been performed. 

We can also understand now why time averaging is essential for quadratic operators. Space averaging of $T$ with a spatial sampling function
$g({\bf x})$ will introduce a factor of its spatial Fourier transform, $\hat{g}({\bf k})$, into each of $A_{i j}$ and $B_{i j}$, if we use a plane wave
mode basis. In particular, now  $B_{i j} \propto  \hat{g}({\bf k_i}+{\bf k_j} )$. However, this factor is not suppressed for pairs of modes where
${\bf k_i} = -{\bf k_j}$. If there is no time averaging, and hence no factors of  $\hat{f}(\omega_i + \omega_j)$, then $\mu_2$ receives a divergent
contribution from such modes.

 The dimensionless measure of the value of the averaged operator $\bar{T}$ is
 \begin{equation}
x = \tau^{p+1} \, \bar{T}\,.
\label{eq:x-def}
\end{equation}
 We seek the probability distribution $P(x)$, and especially its asymptotic form for $x \gg 1$. This form is determined by the behavior of the
 moments $\mu_n$ for large $n$.  The general moment $\mu_n$ may be expressed as a mode sum of an $n$-th degree polynomial in the 
 $A_{i j}$ and $B_{i j}$, and hence in the  Fourier transform, $\hat{f}$. If the asymptotic form in Eq.~\eqref{eq:fasympt} holds, and there is no
 spatial averaging then~\cite{FF2015} 
 \begin{equation}
\mu_n \propto (p\, n/\eta)!
\end{equation}
for large $n$. This rapid growth of the moments leads to a slow decrease in $P(x)$ for large $x$. Specifically,
\begin{equation}
P(x) \sim c_0\, x^b\, {\rm e}^{-a x^c}
\label{eq:P}
\end{equation}
where the constants $c_0$, $b$, $a$, and $c$ depend upon the choice of sampling function.  In particular, with only time averaging, we have
\begin{equation}
c = \frac{\eta}{p}
\label{eq:c-worldline}
\end{equation}
and
\begin{equation}
b = \frac{2-\eta}{p} -(\eta +1) \,.
\label{eq:b}
\end{equation}
The most crucial of these constants is $c$, and its relatively small value describes the slow rate of decrease of $P(x)$.
For example, a stress tensor component such as energy density, with $p=3$, sampled with a temporal sampling function
with $\eta = 1/2$, will have $c = 1/6$, implying a relatively large probability for large fluctuations.

The vacuum is a state with zero total energy, but fluctuations in the energy contained within finite regions of space or time are
possible. We can view these fluctuations as a temporary loan which must be returned within a finite time, possibly through anticorrelated
fluctuations. The timescale for the return of energy to the vacuum is inversely related to the magnitude of the energy. This is 
enforced by the probability distribution, Eq.~\eqref{eq:P}. For a fixed magnitude of the averaged energy density, the dimensionless
variable $x$ increases with increasing $\tau$, causing $P(x)$ to decrease. Thus the probability of finding a given value of the
averaged energy density $ \bar{T}$ decreases with increasing averaging time.

The case of averaging both in time and in space is treated in Ref.~\cite{FF20}, and here we summarize some of the key results.
Now we have a temporal sampling function of width $\tau$ and a spatial function $g({\bf x})$ of characteristic radius $\ell$.
Here we are interested in the case where $\ell \ll \tau$, so we expect a regime where only the temporal sampling is significant.
This arises if $x$, still defined by Eq.~~\eqref{eq:x-def}, is not too large. If $1 \ll x \alt x_*$, where
\begin{equation}
x_* = \left(\frac{2}{a}\right)^{p/\eta} \; \left(\frac{\tau}{\ell}\right)^p \,,
\label{eq:x*}
\end{equation}
then $P(x)$ is approximately given by Eq.~\eqref{eq:P} with $c$ given by Eq.~\eqref{eq:c-worldline}. However, for $x \agt x_*$,
$P(x)$ makes a transition to a regime with the same functional form, but with  $c$ given by
\begin{equation}
c = \eta\,.
\label{eq:c-STA}
\end{equation}
Now $P(x)$ falls more rapidly than in the worldline approximation, where spatial averaging is ignored, so the effect of spatial averaging
is to somewhat decrease the probability of large fluctuations. However, with $\eta < 1$, the decay is still slower than exponential.

The results in Refs~\cite{FF2015,FF20}, which have just been summarized, were obtained by a study of the behavior of the moments.
There is an alternative approach to finding $P(x)$ which involves diagonalization of the operator $\bar{T}$ by a Bogolubov transformation.
This allows the determination of the eigenstates and eigenvalues of this operator. The probability of finding a given eigenvalue in a
measurement is the squared overlap of the corresponding eigenstate with the vacuum, which yields $P(x)$. In practice, this procedure
needs to be performed numerically in a system with a finite number of modes. This was done in Ref.~\cite{SFF18}  for the worldline case
and in Ref.~\cite{WSF21} for the spacetime averaged case, and the results are in good agreement with those of the moments approach.

\subsection{Sampling Functions with Two Time Scales}
\label{sec:2time}

\begin{figure}[htbp]
\includegraphics[scale=0.15]{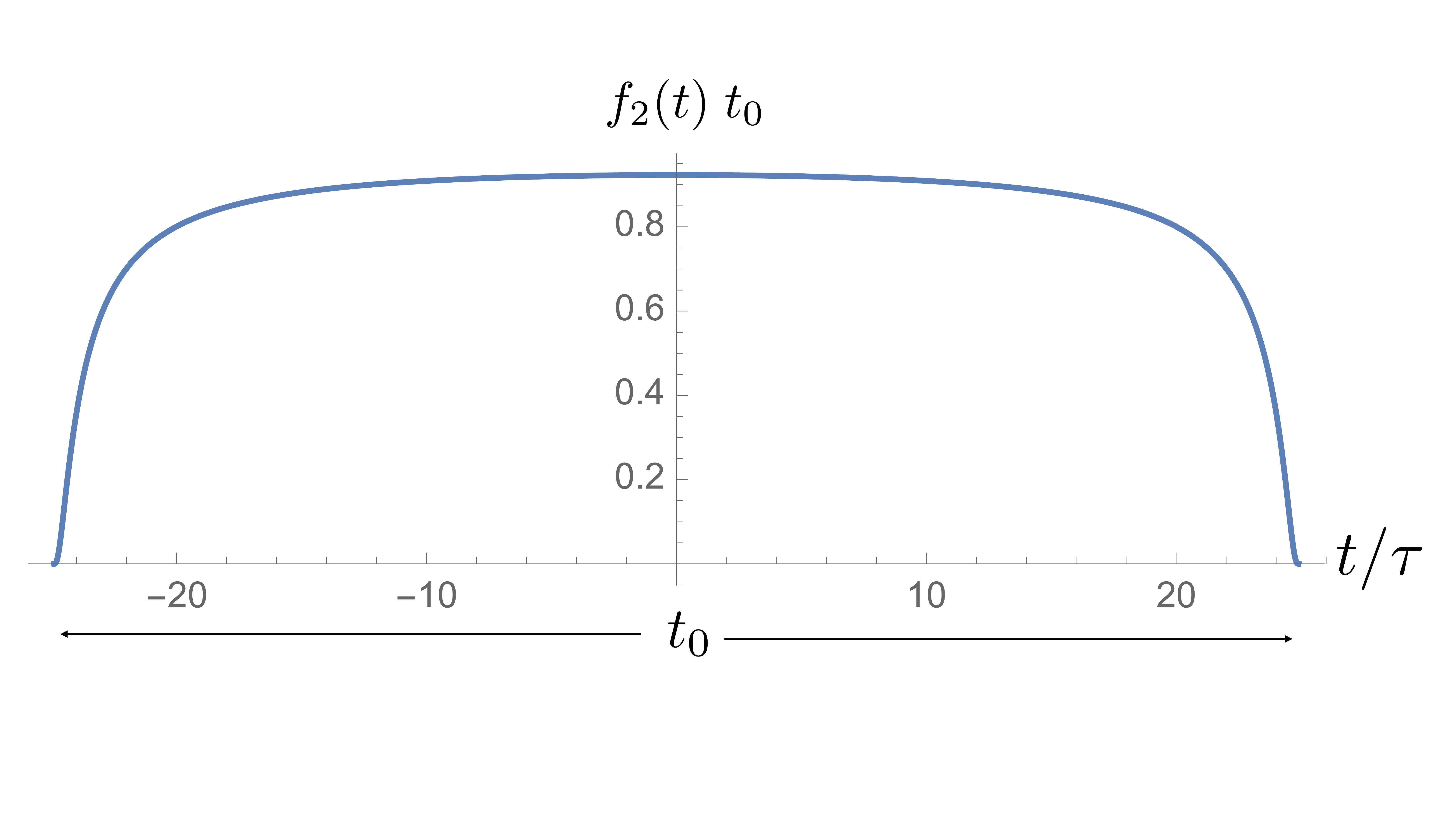}
\caption{A two-time scale sampling function $f_2(t)$, with a duration of $t_0$ and switch-on and switch-off times of the order of $\tau$. Here $t_0 = 50 \, \tau$,
and $f_2(t)$ was constructed using Eqs.~\eqref{eq:F} and \eqref{eq:f2}.}
\label{fig:f2}
\end{figure}

\begin{figure}[htbp]
\includegraphics[scale=0.15]{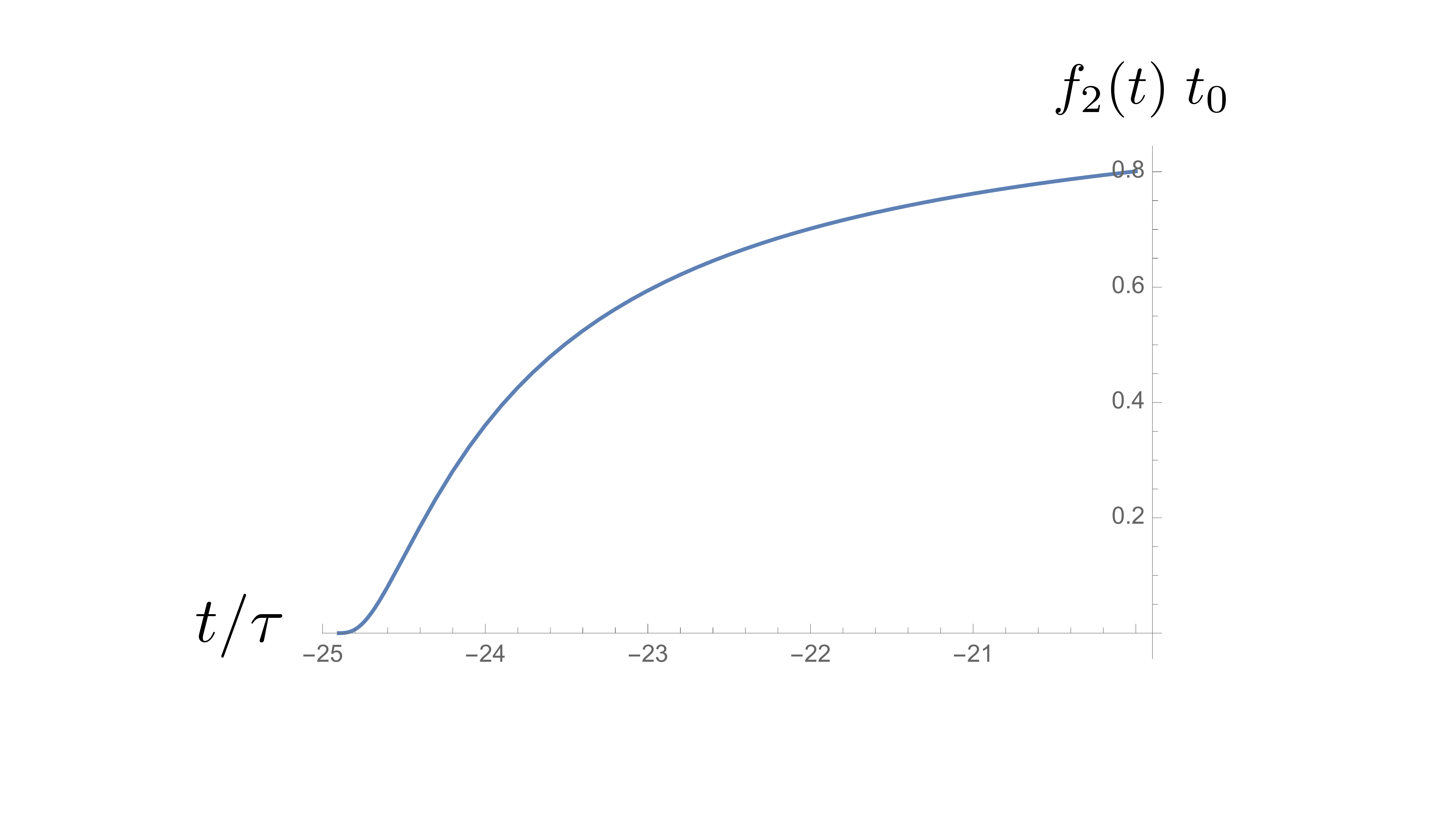}
\caption{Here the switch-on of the function $f_2(t)$ with $t_0 = 50 \, \tau$ is illustrated in detail.}
\label{fig:f2-switch}
\end{figure}

In Sec.~\ref{sec:compact}, we discussed sampling functions of time which depend upon a 
single time scale, $\tau$. Now we consider functions which depend upon two different scales, $\tau$, which is the characteristic rise or
decay time, and a potentially much longer duration, $t_0$. 
Let $F(t)$ be a function which varies from 
values of zero to one on a timescale of the order of $\tau$, so  $F(t) = 0$ for $t<0$ and $F(t) \rightarrow 1$ for $t \agt \tau$. An explicit example is
\begin{equation}
F(t) = {\rm e}^{-\tau/t} \qquad  t > 0\,.
\label{eq:F}
\end{equation}
We can define a two-time scale sampling function by
\begin{equation}
f_2(t) = \frac{C_2}{t_0}\, F(t +t_0/2)\, F(t_0/2 - t) \,. \label{eq:f2}
\end{equation}
This function switches on at $t = - t_0/2$ and off again at $t = t_0/2$, and hence has a total duration of $t_0 \geq 2 \tau$. The rise and fall time is
characterized by $\tau$. If $F(t)$ has the form given in Eq.~\eqref{eq:F}, then $\tau$ is the time required for $F(t)$ to increase from zero to $1/{\rm e}$
of its final value, so the total switch-on or switch-off time for $f_2(t)$ will be a few times  $\tau$. An example of such a function is illustrated in Fig.~\ref{fig:f2}.
The switch-on region of the same function is shown in more detail in Fig.~\ref{fig:f2-switch}.

We can define a single time scale sampling functions, such as those used in Sec.~\ref{sec:compact}, by setting $t_0 = 2\, \tau$ and writing
\begin{equation}
f_1(t) = \frac{C_1}{2 \tau}\, F(t + \tau) \, F(t - \tau) \,. \label{eq:f1}
\end{equation}
Here $C_1$ and $C_2$ are constants defined so that these functions are normalized as in Eq.~\eqref{eq:f-norm}. Note that if $t_0 \gg 2\tau$, then
$C_2 \approx 1$.

The Fourier transforms of these temporal functions, $\hat{F}(\omega)$, $\hat{f_1}(\omega)$, and $\hat{f_2}(\omega)$, are defined as in Eq.~\eqref{eq:Fourier}, 
and are related to one another.  In particular, Eq.~\eqref{eq:f2} and the convolution theorem imply
\begin{equation}
\hat{f_2}(\omega) = \frac{C_2}{ 2 \pi t_0}\, \int^\infty_{-\infty} d\omega_1\, \hat{F}(\omega_1)\, \hat{F}(\omega_1+ \omega) \,{\rm e}^{i \omega_1 t_0} \,.
\label{eq:hat-f2}
\end{equation}
Note that the exponential factor will suppress contributions to the integral from values of $\omega_1 \gg 1/t_0$. Hence, if $\omega \gg 1/t_0$,
we have $\hat{F}(\omega_1+ \omega)  \approx \hat{F}(\omega)$ and
 \begin{equation}
\hat{f_2}(\omega) \approx \frac{C_2}{ 2 \pi t_0}\, \hat{F}(\omega)\, \int^\infty_{-\infty} d\omega_1\, \hat{F}(\omega_1)\,{\rm e}^{i \omega_1 t_0}
= \frac{C_2}{t_0}\, F(t_0)\, \hat{F}(\omega)\,.
\label{eq:hat-f2-asy}
\end{equation}
If we set $t_0 = 2\tau$, then the above result gives the asymptotic form of $\hat{f_1}(\omega)$ for large $\omega$. We see that  $\hat{F}(\omega)$, 
$\hat{f_1}(\omega)$, and $\hat{f_2}(\omega)$ all have the same asymptotic functional forms, given by  Eq.~\eqref{eq:fasympt} with the same values
of $\beta$ and $\eta$, but different values of $\gamma$.

Because $f_1(t)$ and $f_2(t)$ are normalized sampling functions which satisfy Eq.~\eqref{eq:f-norm},   Eq.~\eqref{eq:Fourier} implies that 
$\hat{f_1}(0) =\hat{f_2}(0) =1$. However, $F(t)$ is not normalizable, so we expect $\hat{F}(\omega) \rightarrow \infty$ as $\omega \rightarrow 0$.
Because $F(t)$ is a dimensionless function, $\hat{F}(\omega)$ has dimensions of time and is hence proportional to $\tau$. For $\omega \gg 1/\tau$,
we expect
\begin{equation}
\hat{F}(\omega) \approx \gamma_0 \, \tau \,  {\rm e}^{- \beta |\omega \tau|^\eta} \, ,
\end{equation}
where $\gamma_0$ is a constant of order one. If $t_0 \gg \tau$, where $C_2 \approx 1$ and $F(t_0) \approx 1$, Eq.~\eqref{eq:hat-f2-asy} becomes  
\begin{equation}
\hat{f_2}(\omega) \approx \gamma_0 \, \frac{\tau}{t_0}\,  {\rm e}^{- \beta |\omega \tau|^\eta} \, .
\label{eq:hat-f2-asy2}
\end{equation}

\subsection{The Parameters $\eta$ and $\beta$}
\label{sec:eta-beta}

The rate of decay of the Fourier transform of a sampling function for large $\omega$ is given by Eq.~\eqref{eq:fasympt}, which effectively defines the
parameters $\eta$ and $\beta$. Both of these parameters play important roles in our discussion, as the probability distribution  describing
large fluctuations is sensitive to both. The greater sensitivity is to the value of $\eta$, which appears in the power of $\omega$ in the exponential
in $\hat{f}(\omega)$, and in the power of  $x$ in $P(x)$, so smaller values of $\eta$ are linked to more slowly decreasing $\hat{f}(\omega)$ and to
a greater probability of large fluctuations. The same is true to a lesser degree of the value of $\beta$, which appears as a coefficient inside the
exponential functions in both $\hat{f}(\omega)$ and $P(x)$. 

The value of $\eta$ is associated with the functional form of $f(t)$ near the switch-on time, which we take here to be $t=0$. The class of compactly
supported functions, whose Fourier transforms have the asymptotic form in Eq.~\eqref{eq:fasympt}, have the switch-on behavior
\begin{equation}
f(t) \sim t^{-\mu} \, {\rm e}^{- w\, t^{-\nu}}\,, \quad t \rightarrow 0^+ \,,
\label{eq:small-t}
\end{equation}
 where $\nu = \eta/(1-\eta)$, $w$, and $\mu$ are constants determined by $\eta$. [See Eqs. (50)-(54) in Ref.~\cite{FF2015}]. In particular,
 $\eta = 1/2$ is associated with $\nu =1$, as illustrated in Eq.~\eqref{eq:F}. As noted above, the functions $F(t)$, $f_1(t)$, and $f_2(t)$, related
 by Eq.~\eqref{eq:f2}, or similar relations, all have the same values for  $\eta$ and $\beta$.
Recall that the electrical model in  Ref.~\cite{FF2015} creates a switch-on described by a function with $\eta = 1/2$. The value of $\beta$
can vary in this model, but is determined by the parameters of the specific circuit. 

We can better understand the role of the parameter $\beta$ from the study of a class of functions discussed by Johnson~\cite{Johnson} of the form
\begin{equation}
f_J(t) =  C\, {\rm e}^{- \beta^2\, [1 - (t/\tau)^2]^{1-a}} \,,
\label{eq:fJ}
\end{equation}
 where $b > 0$, $a > 1$,  $-\tau < t < \tau$, and $C$ is a normalization constant.  Note that the switching behavior is similar to that in Eq.~\eqref{eq:small-t}, 
 with $\nu = a - 1$, but without
 the factor of $ t^{-\mu}$. Johnson evaluates the asymptotic form for the Fourier transform $\hat{f_j}(\omega)$ using a saddle point approximation, and finds
 for the case $a = 2$ that
 \begin{equation}
\hat{f_J}(\omega) \propto \omega^{-3/4} \,  {\rm e}^{-\beta\,  \sqrt{ \tau \,\omega}} \,, \quad \omega \gg 1/\tau \,.
\label{eq:fJhat}
\end{equation}
This shows that $a=2$ corresponds to $\eta = 1/2$, and in this case,  the parameter $\beta$ in Eq.~\eqref{eq:fJ} is the same as that in Eq.~\eqref{eq:fasympt}.
Note that $\hat{f_J}(\omega)$ contains a factor of $\omega^{-3/4}$ which does not appear in  Eq.~\eqref{eq:fasympt}, and $f_J(t)$ lacks the factor of  $t^{-\mu}$
which appears in Eq.~\eqref {eq:small-t}, which show that $f_J(t)$ and  $f(t)$ are somewhat different functions. Nonetheless, they share the same exponential 
factor in their Fourier transforms. 

\begin{figure}[htbp]
\includegraphics[scale=0.15]{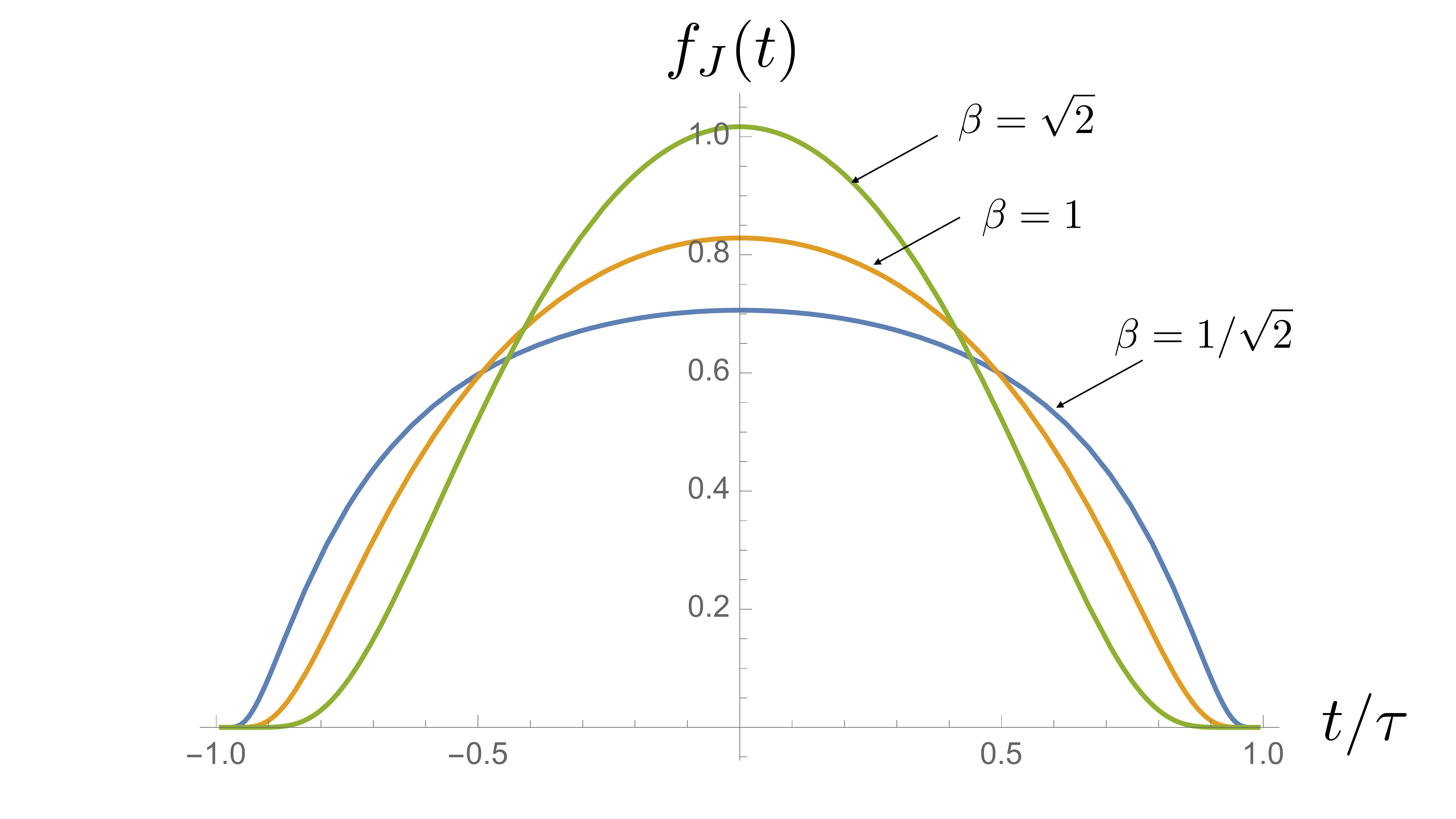}
\caption{The normalized function $f_J(t)$ is plotted for the case $a=2$, or $\eta = 1/2$, for three values of $\beta$}
\label{fig:fJ-beta}
\end{figure}

Here we assume that the $\beta$-dependence of $f_J(t)$ is a reasonable guide to understanding that of  $f(t)$. In Fig.~\ref{fig:fJ-beta}, $f_J(t)$ is plotted for the case $a=2$  for
three values of $\beta$. We can see that smaller $\beta$ leads to more rapid switch-on and switch-off, as one might expect from the increased contribution of
high frequency Fourier components as $\beta$ decreases.

\subsection{Vacuum Radiation Pressure Fluctuations on Atoms: Probability Distribution}
\label{sec:atoms}

Vacuum radiation pressure fluctuations on a perfect mirror could in principle be computed from those of the appropriate components of the 
electromagnetic stress tensor. However, in the present paper we are concerned with atoms, which scatter longer wavelengths of light
by Rayleigh scattering. The vacuum radiation pressure fluctuations on such polarizable particles were treated in Sec.~V of Ref.~\cite{HF17}, 
and the results will be summarized here. The Rayleigh cross section for the scattering of light with angular frequency $\omega$ from a
particle with a static polarizability $\alpha$ is
\begin{equation}
\sigma_R = \frac{\alpha^2}{6 \, \pi} \, \omega^4 \,.
\end{equation}
The momentum flux in the incident beam is given by $\mathbf{E} \times \mathbf{B}$, where $\mathbf{E}$ and $\mathbf{B}$ are the quantized
electric and magnetic field operators. The net momentum carried by the scattered radiation vanishes, so the the force in the $z$-direction on 
the particle may be written as
\begin{equation}
F^z = \sigma_R\, (\mathbf{E} \times \mathbf{B})^z =  \frac{\alpha^2}{6 \, \pi} \; R^z\,,
\end{equation}
where 
\begin{equation}
R^z =  (\mathbf{\ddot{E}} \times \mathbf{\ddot{B}})^z
\end{equation}
may be viewed as the radiation pressure operator for polarizable particles.
Note that the factor of $\omega^4$ in the Rayleigh cross section has resulted in two time derivatives on each of the electromagnetic fields.
As a consequence, $R^z$ is an operator with $p=7$.

We may average $F^z$ in time by letting $\alpha = \alpha(t)$ and integrating along the worldline of the atom. A specific model for the
origin of this time dependence will be presented in the next section. We take the time dependence to be proportional to the two-scale
sampling function, $f_2(t)$, discussed in Sec.~\ref{sec:2time} with $t_0 \gg \tau$. Here we set $\alpha^2(t) \propto \alpha_0^2 \, f_2(t)$, where
$\alpha_0^2$ is the time average of   $\alpha^2(t)$ 
The averaged force now becomes $\bar{F^z} = \alpha_0^2\, \bar{R^z}/(6 \pi)$, where 
\begin{equation}
\bar{R^z} = \int f_2(t) \,  R^z(t)  \,  dt \,,
\label{eq:ave-pressure}
\end{equation}
The probability distribution for large fluctuations of $\bar{R^z}$, and hence of $\bar{F^z}$, is given by
Eqs.~\eqref{eq:P}, \eqref{eq:c-worldline}, and \eqref{eq:b}, with  $p=7$ and $\eta$ determined by $\alpha^2(t)$.

Let $x = \tau^8 \,\bar{R^z}$ be the dimensionless measure of the quantum radiation pressure fluctuations. For $1 \ll x < x_*$,
its probability distribution has the asymptotic form in Eq.~\eqref{eq:P}, with $c = \eta/7$ and $b= -(8 \eta +5)/7$ from
Eqs.~\eqref{eq:c-worldline} and \eqref{eq:b}. The remaining constants in Eq.~\eqref{eq:P} may be found from Eqs.~(100)
and (101) in Ref.~\cite{FF2015} with $\alpha \rightarrow \eta$, $B_0 =4$, and  $B= 1/(6 \pi^2)$. (See Eq.~(35) in Ref.~\cite{HF17}.)
In addition we now set $\gamma = \gamma_0\, \tau/t_0$ and $f(0) \rightarrow \tau\, f_2(0) \approx  \tau/t_0$. The last factor of $\tau$
arises because  Ref.~\cite{FF2015} uses units where $\tau = 1$, so $f(0)$ would be $\tau \,f(0)$ in general units. Thus we find
\begin{equation}
a = 2 \beta \, \left(\frac{\tau}{ 3\pi \, t_0} \right)^{-\eta/7}   \,,
\label{eq:a}
\end{equation}
and 
\begin{equation}
c_0 = \frac{\gamma_0^2}{128 \, \pi^2 \, (\beta \, \eta)^8} \;  \left(\frac{\tau}{ 3\pi \, t_0} \right)^{2(4 \eta -7 )/4} \,.
\label{eq:c0}
\end{equation}
If we set $\eta = 1/2$ then the constants in Eq.~\eqref{eq:P} become
\begin{equation}
c = \frac{1}{14}\,, \qquad b= -\frac{9}{7} \,, \qquad  a = 2 \beta \, \left(\frac{\tau}{ 3\pi \, t_0} \right)^{-\frac{1}{14}}   \,, 
\label{eq:cba}
\end{equation}
and
\begin{equation}
c_0 = \frac{2\, \gamma_0^2}{\pi^2 \, \beta^8} \;  \left(\frac{\tau}{ 3\pi \, t_0} \right)^\frac{2}{7} \,.
\label{eq:c0-2}
\end{equation}
Now the asymptotic probability density has the form
\begin{equation}
P(x) = c_0 \, x^{\frac{9}{7}} \,{\rm e}^{-a\, x^\frac{1}{14}} \,.
\label{eq:P2}
\end{equation}

If $p=7$ and $\eta = 1/2$, the upper limit for the validity of the worldline approximation, given by Eq.~\eqref{eq:x*}
and the value of $a$ in Eq.~\eqref{eq:cba},    becomes
\begin{equation}
x_*  = \beta^{-14}\,  \left(\frac{\tau}{ 3\pi \, t_0} \right) \, \left(\frac{\tau}{ \ell} \right)^7 \,. 
\label{eq:x*-2}
\end{equation}

It will be useful to have a measure of the actual probability of a fluctuation of the order of magnitude of a given value of $x$.
For this purpose, we define the quantity
\begin{equation}
{\cal P}(x) = \int_{\frac{1}{2} x}^x P(y) \, dy \,,
\end{equation}
which is probability of finding an outcome in the interval $\frac{1}{2} x \leq y \leq x$. Use the form of $P$ in Eq.~\eqref{eq:P2} and
define $u = y^\frac{1}{14}$ to write
 \begin{equation}
{\cal P}(x) = 14 c_0 \int_{u_1}^{u_2}  u^{-5} \, {\rm e}^{-a\, u} \, du \approx 14\, c_0 \, {u_2}^{-5}\, {\rm e}^{-a\, u} \, \Delta u \,,
\end{equation}
where $u_1 = (\frac{1}{2} x)^\frac{1}{14}$,  $u_2 = x^\frac{1}{14}$ , and $\Delta u = u_2 -u_1 \approx 0.0483\, u_2$. This leads to
\begin{equation}
{\cal P}(x) \approx 0.676 \,x\, P(x) = 0.676\, c_0 \, x^{-\frac{2}{7}}\, {\rm e}^{-a\, x^\frac{1}{14}} \,.
\label{eq:cumm-P}
\end{equation}  

Recall that in Sec.~\ref{sec:stress} we discussed how the probability distribution enforces repayment of an energy loan from the vacuum.
Similarly, a radiation pressure fluctuation on an atom amounts to a temporary loan of linear momentum. Suppose that we hold $\tau$
fixed but increase $t_0$. In this case, $x = \tau^8 \,\bar{R^z}$ does not change. However, if $\eta = 1/2$, then from Eqs.~\eqref{eq:a} and \eqref{eq:c0-2},
we have $a \propto t_0^\frac{1}{14}$ and  $c_0 \propto  t_0^{-\frac{2}{7}}$. Both of these effects act to decrease ${\cal P}(x)$ as  $t_0$
increases.

\section{Switched Polarizability and Velocity Fluctuations of Atoms}
\label{sec:v-flucts}

In this section, we present a model for the origin and effects of a time-dependent polarizability of an atom.

\subsection{Switching with Rydberg Atoms}
\label{sec:Rydberg}
 Rydberg atoms are hydrogen-like atoms in highly excited states, and have been extensively investigated in recent years.
 For a review, see for example, Gallagher~\cite{Gallagher}. If $n$ is the principal quantum number of the excited state,
 several physical properties of the Rydberg atom scale as powers of $n$. These include the mean radius, $r_0 \propto n^2$,
 the polarizability, $\alpha \propto n^7$, and the radiative lifetime, $\tau_{rad} \propto n^3$. See Table 2 in Ref.~\cite{Gallagher}. 
 This means that the polarizability, can increase many orders of magnitude from its value in the ground state when the atom
 is excited, maintain this value for a time as long as $\tau_{rad}$, and then return to a much smaller value when the atom
 is de-excited. The excitation can be produced by a short laser pulse, which at optical frequencies can have a duration of the
 order of $1 {\rm fs}$. This pulse is assumed to be described by a compactly supported function of time, and a frequency spectrum 
 of the form of Eq.~\eqref{eq:fasympt}.  The de-excitation can be induced by a second such pulse, which causes the polarizability to decrease
 rapidly either by stimulated emission or by ionization. In this case, the polarizability $\alpha(t)$ may have a time-dependence
 of the form illustrated in Fig.~\ref{fig:f2}. 
 
 We may take
\begin{equation}
\alpha^2(t) = \alpha_0^2 \, t_0 \, f_2(t) \,,
\end{equation}
where $\alpha_0$ is the polarizability in the excited state and $f_2(t)$ is a two-scale sampling function of the type discussed 
in Sec.~\ref{sec:2time}. The factor of $t_0$ arises because $f_2(t)$ satisfies Eq.~\eqref{eq:f-norm}. Recall that $f_2(t)$, and
hence $\alpha^2(t)$, increases and later decreases on a time scale $\tau$, the duration of the exciting and de-exciting pulses.
Here we take $t_0 \gg \tau$ to be the approximate temporal duration of the excited state. Because  $\alpha^2(t)$ increases and
then decreases by many orders of magnitude, we may take its initial and final values to be zero.

This time dependence leads to vacuum radiation pressure fluctuations, with the averaged radiation pressure in the $z$-direction
being given by $\bar{R^z}$ defined in Eq.~\eqref{eq:ave-pressure}. Note that here the $z$-direction is randomly selected and
need not be correlated with the direction of either laser pulse. If the excited state is approximately spherically symmetric, then
we expect vacuum radiation pressure fluctuations to be equally likely in any direction. In this case,  $\langle \bar{R^z} \rangle =0$,
and all other odd moments will also vanish. However, the variance,  $\langle (\bar{R^z})^2 \rangle$ and other even moments are
nonzero. This leads to a symmetric probability probability distribution, $P(x) = P(-x)$, as fluctuations of either sign and in any direction
are equally probable. The asymptotic form of $P(x)$ for large $x$ can be taken to be given by Eq.~\eqref{eq:P}, where the parameters
such as $\beta$ and $\eta$ are determined by the Fourier transforms of the pulses which cause the switching of the polarizability.
In particular, if these pulses have $\eta =1/2$, then $P(x)$ takes the form in Eq.~\eqref{eq:P2} and falls very slowly.

\subsection{Velocity Fluctuations: Numerical estimates}
\label{sec:estimates}

Now we wish to estimate the magnitude and probability of the velocity fluctuations of the atom due to vacuum radiation pressure fluctuations.
For simplicity, we assume that the fluctuations may be describe in the worldline approximation and with $\eta =1/2$, so we may use Eq.~\eqref{eq:P2}
with $x \alt x_*$, with  $x_*$ given by Eq.~\eqref{eq:x*-2}. We take the probability of a fluctuation with $x \approx x_*$ to be ${\cal P}(x_*)$, given by
Eq.~\eqref{eq:cumm-P}. Such a fluctuation produces a radiation pressure of order $x_*/\tau^8$ and a  force which magnitude is of the order of
\begin{equation}
\bar{F} \approx \frac{\alpha_0^2}{ 6 \pi \, \tau^8}  \, x_*  
\end{equation}
and a duration of the order of $t_0$. The characteristic recoil speed will be of order
\begin{equation}
\bar{v} \approx \frac{ \bar{F} \, t_0}{m}  \approx  \frac{\alpha_0^2 \, t_0}{ 6 \pi \,m \, \tau^8}  \, x_*    
\end{equation}
where $m$ is the mass of the atom. 

We may use Eq.~\eqref{eq:x*-2} and set $\ell \approx r_0$, the characteristic size of the Rydberg atom, to write
\begin{equation}
\bar{v} \approx \frac{\alpha_0^2}{ 18 \, \pi^2 \,m\, \beta^{14}\, r_0^7}\,.
\label{eq:v}
\end{equation}
There are two remarkable feature of this expression: first, the dependence upon both $\tau$ and $t_0$ have cancelled. Second, the result is
very sensitive to the parameter $\beta$, which appears in the asymptotic form of the Fourier transform of the sampling function, 
Eq.~\eqref{eq:fasympt}, and hence in the frequency spectrum of the exciting and de-exciting laser pulses.
The meaning of this parameter was discussed in Sec.~\ref{sec:eta-beta}. However, we should note that $\beta$ makes its
appearance in $\bar{v}$ through the $\beta$-dependence of $x_*$, and hence linked with our restriction to using the worldline approximation.
We may express Eq.~\eqref{eq:v} as
\begin{equation}
\bar{v} \approx \frac{2.1 \times 10^{-8}}{\beta^{14}} \, \left(\frac{1 {\rm u}}{m} \right) \,  \left(\frac{\alpha_0}{n^7 \, a_0^3} \right)^2  \, \left(\frac{n^2 \, a_0 }{r_0} \right)^7\,, 
\end{equation}
where $a_0$ is the Bohr radius of hydrogen.
Note our estimate for $\bar{v}$ is independent of $n$, and  that we expect $\alpha_0 \approx n^7 \, a_0^3$ and $r_0 \approx n^2 \, a_0$, in which case the last two factors in
the above expression will be of order one.
Recall that we are working in units with the speed of light set to one, so $\bar{v} = 2.1 \times 10^{-8} = 6.3\,{\rm m/s}$. Note the very strong dependence upon the value of
$\beta$. For example, if $\beta = 1/\sqrt{2}$, and the other factors in Eq.~\eqref{eq:v} are of order one, we would obtain the estimate  
$\bar{v} \approx   2.7 \times 10^{-6} \approx   800 \,{\rm m/s}$.

We can compare these estimates with the expected atomic speeds due either to thermal motion, or the recoil from photon absorption or emission. At absolute temperature
$T$, the  root-mean-square thermal speed is 
\begin{equation}
v_T = \left( \frac{3\, k_B\, T}{m} \right)^\frac{1}{2} \approx 5 \times 10^{-10} \, \left(\frac{T}{1 \mu {\rm K}} \right)^\frac{1}{2} \; \left(\frac{1 {\rm u}}{m} \right)^\frac{1}{2} \,. 
\end{equation}
Thus at sufficiently low temperatures, the thermal motion can be smaller than the effects of vacuum radiation pressure fluctuations. If an atom at rest either absorbs or emits
a photon  with energy $E_\gamma$, the magnitude of the recoil momentum of the atom will equal the photon's momentum, resulting in a recoil velocity of magnitude
\begin{equation}
v_R \approx 10^{-9} \, \left( \frac{E_\gamma}{ 1 {\rm eV}}  \right) \, \left(\frac{1 {\rm u}}{m} \right)\,.
\end{equation}
This can also be smaller than the effects of vacuum radiation pressure fluctuations, and is also correlated with the photon's direction.

\begin{figure}[htbp]
\includegraphics[scale=0.15]{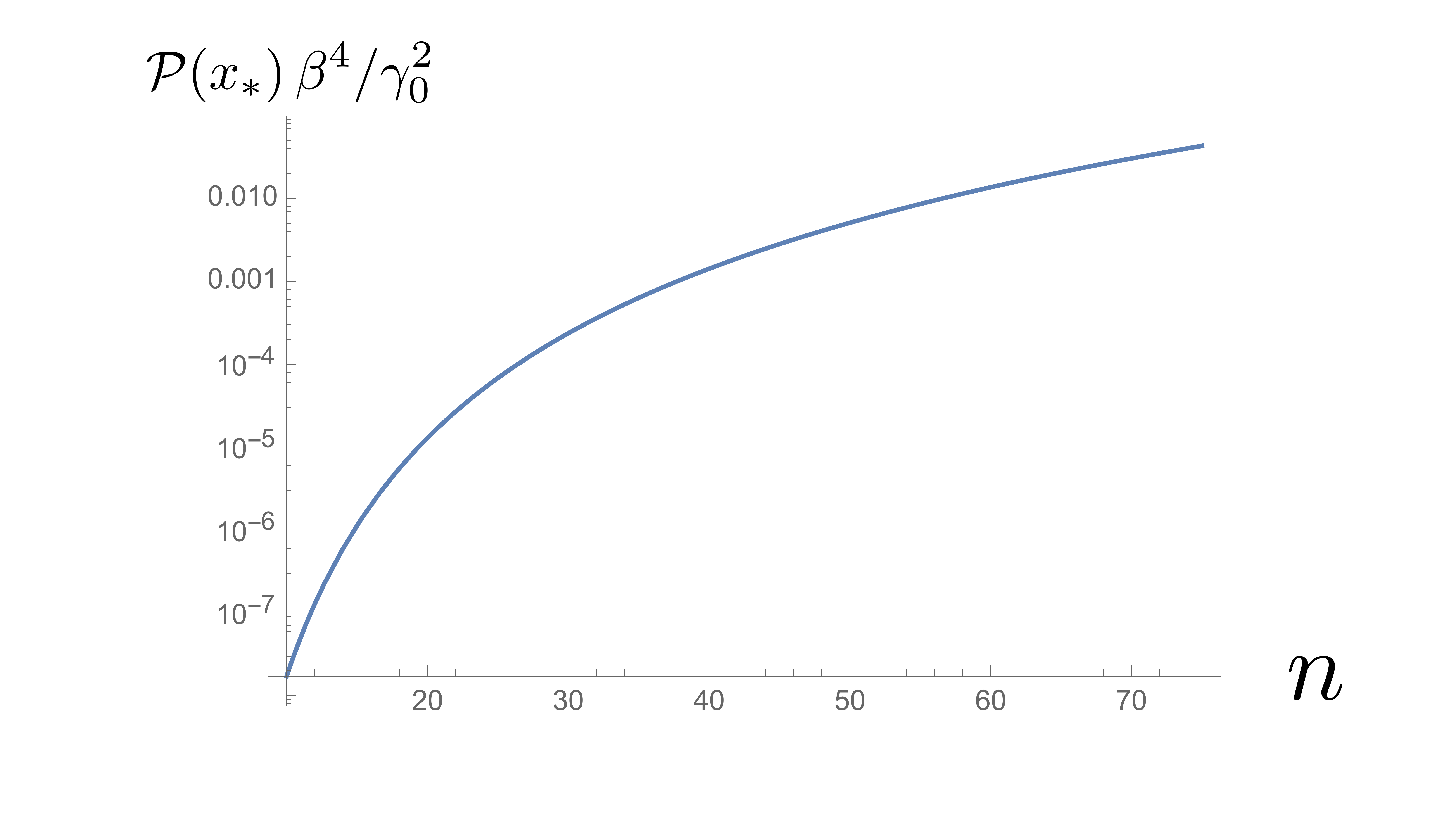}
\caption{The estimate of the probability of a radiation pressure fluctuation in a given measurement is plotted as a function of the principal quantum number $n$. Here we have set
$\tau = 1 {\rm fs}$.}
\label{fig:Px}
\end{figure}

Let us now turn to estimates of the probability ${\cal P}(x_*)$ for the fluctuations in question.  Equations~\eqref{eq:cba}, \eqref{eq:c0-2}, \eqref{eq:x*-2}, 
and \eqref{eq:cumm-P} lead to
\begin{equation}
{\cal P}(x_*) \approx 0.14 \, \gamma_0^2 \, \beta^{-4} \, \left(\frac{r_0}{\tau} \right) \,  {\rm e}^{-2\, \sqrt{\tau/r_0}} \,.
\label{eq:Px*}
\end{equation}
This probability also increases as $\beta$ decreases, although not so strongly as does $\bar{v}$. Recall that $r_0 \approx a_0\, n^2$. If we set 
$\tau = 1 \,{\rm fs} \approx 3 \times 10^{-5}\, {\rm cm}$, then
\begin{equation}
\frac{\tau}{r_0} \approx \frac{6000}{n^2} \approx \left(\frac{77}{n}\right)^2\,.
\end{equation}
 Recall that for the worldline approximation to have any range of validity, we need $n \alt 77$. In Fig.~\ref{fig:Px}, the value of the expression for ${\cal P}(x_*)$ given
 by Eq.~\eqref{eq:Px*} is plotted as a function of $n$.
 If  both $\beta$ and $\gamma_0$ are of order one, then we can have ${\cal P}(x_*)$  of the order of $1\%$ while $\tau > r_0$.
 
We have presented estimates both for the mean speed, $\bar{v}$, and the probability, ${\cal P}(x_*)$, of a vacuum radiation pressure fluctuation capable of producing this mean
speed. Both of these estimates assumed the worldline approximation, in which the radiation pressure operator is averaged in time only, and are hence restricted to considering
fluctuations for which $x \alt x_*$. This restriction is probably responsible for the strong dependence of ${\cal P}(x_*)$, and especially of $\bar{v}$, on the parameter $\beta$. 
This may be seen from the sensitivity of  $x_*$ to the value of $\beta$ exhibited in Eq.~\eqref{eq:x*-2}. Thus it is reasonable to expect that including spatial averaging, which
will allow consideration of larger fluctuations, may reduce the dependence of the results upon $\beta$. This will be a topic for future work.

\section{Summary}
\label{sec:final}
 
This paper has dealt with vacuum radiation pressure fluctuations on Rydberg atoms. Such fluctuations are governed by a probability distribution which decreases very slowly
and is very sensitive to the details of how the radiation pressure  is measured. The relevant quantum operator is quadratic in second time derivatives of the electromagnetic
field. Like other stress tensor-type operators for the quantized electromagnetic field, it must be averaged in time before its fluctuations become finite. This averaging describes
the measurement process of the radiation pressure. If the measurement is to begin and end at finite times, then the temporal averaging function must have compact support,
meaning that it is a non-analytic, but $C^\infty$, function which vanishes outside of a finite time interval. Here a specific model for the measurement has been proposed, in
which an atom is excited from its ground state to a highly excited state, forming a Rydberg atom with greatly increased polarizability. The excitation is assumed to be caused
by a short laser pulse of finite duration, after which the excited state persists for a time long compared to the excitation time. A second short pulse ends the excitation by either
ionization or stimulated emission. 

The effect of the resulting time-dependent atomic polarizability is to measure the vacuum radiation pressure on the atom, which in turn causes a temporary recoil of the atom
that may be large enough to be observed. The atom in effect borrows some linear momentum from the quantum vacuum for a finite time. The probability distribution for large
radiation pressure fluctuations decreases remarkably slowly at a rate determined by the pulse frequency spectrum, such as an exponential of the $\frac{1}{14}$ power of the
pressure fluctuation. It is this slow rate of decrease which allows the possibility of observable recoil with non-negligible probability.

The treatment in this paper has assumed a worldline approximation in which the radiation pressure  operator is averaged in time, but not in space. This approximation
simplifies the analysis, but limits the magnitude of the fluctuations which may be considered. An expanded treatment including spatial averaging, using either the analytic
methods of Refs.~\cite{FF2015,FF20}, or the numerical methods of Refs.~\cite{SFF18,WSF21} will be a topic for future research.

\begin{acknowledgments} 
I would like to thank Haiyun Huang, Enrico Schiappacasse, and Peter Wu for useful conversations.
This work was supported in part  by the National Science Foundation under Grant PHY-1912545.
\end{acknowledgments}

\end{document}